\documentstyle[12pt]{article}

\newcommand{\ket}[1]{\left | \, #1 \right \rangle}

\begin{document}

\bigskip
\bigskip
\begin{center}
{\large Quantum Entanglement and the Nonexistence of Superluminal Signals \\
}\bigskip

Michael D. Westmoreland,$^{(1)}$ Benjamin Schumacher$^{(2)}$
\bigskip \\

{\small{\sl

$^{(1)}$Department of Mathematical Sciences, Denison University,
	Granville, OH  43023 USA \\
$^{(2)}$Department of Physics, Kenyon College, Gambier, OH 43022 USA }}
\bigskip
\end{center}
\bigskip

\subsection*{\centering Abstract}
{This paper shows that ordinary quantum mechanics is not consistent 
with the superluminal transmission of classical information.}
\vfill

PACS numbers: 03.65.Bz, 42.50.Dv, 89.70.+c\vfill

\pagebreak

It is well-known that the statistical correlations between entangled
quantum systems (such as a pair of spins in a singlet state) cannot be used to send classical messages faster than light \cite{superlum}.
In this paper we show a stronger result:  that {\em any} means of sending superluminal
signals would violate the laws of quantum mechanics.  Our proof is constructed from
three elements:  the ``no-cloning'' theorem of quantum mechanics \cite{no-cloning},
quantum teleportation \cite{teleportation}, and the relativity of simultaneity.  

The ``no-cloning'' theorem, as proved by Wootters and Zurek \cite{no-cloning}, is
based on the linearity of quantum mechanics.  
Suppose a scheme exists by which we could copy arbitrary states from a quantum system $X$ 
into a similar system $Y$.  This may involve an interaction with an apparatus $M$.
In copying, the original state of $X$ is undisturbed and $Y$ is 
brought into a state identical to the original state; that is, the combined system 
evolves according to
\begin{eqnarray} 
\ket{a_X,0_Y,0_M} \rightarrow \ket{a_X,a_Y,\psi_{a M}}.  
\end{eqnarray}
Here $\ket{0_Y}$ is some standard ``null'' state of $Y$
and $\ket{0_{M}}$ is the initial state of the cloning apparatus.
After the cloning process, the state of $M$ is $\ket{\psi_{a M}}$,
which may depend on the input state $\ket{a_{X}}$.

Consider the action of our copying scheme on the input state 
$\ket{c_X} \ = \ \alpha \ket{a_X} + \beta \ket{b_X}$, 
where $\ket{a_X}$ and $\ket{b_X}$ are distinct states of $X$.
If the final state of $Y$ is to be a faithful copy, then 
$\ket{c_Y} \ = \ \alpha \ket{a_Y} + \beta \ket{b_Y}$.
From general considerations of quantum mechanics, 
we know that the evolution of the system is {\em linear}, so that
\begin{eqnarray} 
\ket{c_X,0_Y,0_{M}} 
& = & \alpha \ket{a_X,0_Y,0_{M}} + \beta \ket{b_X,0_Y,0_{M}} \\
& \rightarrow & \alpha \ket{a_X,a_Y,\psi_{aM}} + \beta \ket{b_X,b_Y,\psi_{bM}} \\
& \neq & \ket{c_X,c_Y,\psi_{cM}}, 
\end{eqnarray}
since $\ket{c_X,c_Y} \ = \ \alpha^{2} \ket{a_X,a_Y} + \alpha \beta \ket{a_X,b_Y} 
+ \beta \alpha \ket{b_X,a_Y} + \beta^{2} \ket{b_X,b_Y}$. 
Thus, if two distinct states can be copied faithfully by some scheme, 
a superposition of them cannot be.

We note that the ``no-cloning'' theorem in fact holds for the most general
sort of quantum evolution described by a completely positive map on density
operators, since all such maps can be modeled by unitary (and thus linear)
evolution on a larger system (a fact reviewed in \cite{sendentang}).  
In particular, cloning is still impossible even if we allow 
measurements and manipulations of the systems based on the outcomes 
of measurements.

We will show that the existence of superluminal signals implies the existence 
of a cloning scheme that uses quantum teleportation \cite{teleportation}. 
In quantum teleportation, a sender Alice conveys an arbitrary state $\ket{\phi_{C}}$ 
of a qubit $C$ to a receiver Bob.  This is done using an entangled pair of qubits 
($A$ and $B$) already
shared by Alice and Bob, together with a classical message transmitted from Alice to Bob.
We may imagine that the entangled qubits are initially in the singlet state 
$\ket{\Psi_{AB}^{-}} = 
\frac{1}{\sqrt{2}}(\ket{\uparrow_A,\downarrow_B} - \ket{\downarrow_A,\uparrow_B})$.
Alice measures a joint observable on $C$ and $A$ whose eigenstates are the so-called
Bell states 
$\ket{\Phi_{AC}^{\pm}} = 
\frac{1}{\sqrt{2}}(\ket{\uparrow_A,\uparrow_C} \pm \ket{\downarrow_A,\downarrow_C})$
and 
$\ket{\Psi^{\pm}_{AC}} = \frac{1}{\sqrt{2}}(\ket{\uparrow_A,\downarrow_C} \pm \ket{\downarrow_A,\uparrow_C})$.
(This measurement tells Alice nothing about the input state $\ket{\phi_{C}}$.)
The measurement result can be sent to Bob by a classical message of two bits.
It turns out that this information is precisely enough to allow Bob to choose a 
unitary transformation for his qubit $B$ that will leave it in the state $\ket{\phi_{B}}$,
exactly the same as the input qubit state.  Thus, by using one entangled pair of qubits
and sending a two-bit classical message, Alice can transfer her unknown state to Bob.
The choice of the singlet state $\ket{\Psi^{-}_{AB}}$ is not essential, and any
``maximally entangled'' state $\ket{\Psi_{AB}}$ of the qubits $A$ and $B$ 
can be used for teleportation.

Assume that Alice and Bob share an entangled pair of qubits and are capable 
of conveying classical information superluminally by some unspecified means.
We can imagine that Alice and Bob are separated by a large distance.
We consider the teleportation of $\phi$ to consist of two events: \\
\begin{itemize}
 \item{I.} Alice's joint measurement of qubits $A$ and $C$ and the 
sending of the classical message. 
 \item{II.} Bob's receipt of the message and his application of the appropriate
unitary transformation to qubit $B$.
\end{itemize}
The original state of $C$ is annihilated at event I, and the final (identical) 
state of $B$ is created at event II.
(By ``large distance'' we mean that the light travel time between Alice and Bob 
is big enough to think of I and II as spacetime points---i.e., ``events''.)

If the classical message is conveyed superluminally, there is a spacelike separation 
between events I and II \cite{relativity}.  
Therefore, there is some frame of reference in which 
II {\em precedes} I in time.  
Consider such a frame of reference, in which the (possibly transformed)
initial state of qubit $C$ is $\ket{\phi'_{C}}$.
During the time interval between events II and I, the qubits $C$ and
$B$ are in the state $\ket{\phi'_{C},\phi'_{B}}$---a state in which $C$ has been ``cloned''
into $B$.  By the no-cloning theorem, this cannot occur at any time.  Hence, either 
quantum mechanics is wrong in this frame of reference
or there can be no superluminal carriers of classical information.

It might be argued that this is not ``real'' cloning, since the original state $\ket{\phi'_{C}}$
is fated to be destroyed before the end of the teleportation process.  Such cloning could
thus not be used, for example, to improve the distinguishability of various input $C$-states.
However, suppose that the choice of input state $\ket{\phi'_{C}}$ is known to both
Alice and Bob.  Immediately before event I, Alice could perform a measurement testing
to see whether qubit $C$ is in state $\ket{\phi'_{C}}$, while Bob could perform
a simultaneous measurement immediately after event II to test whether qubit $B$ is
in the state $\ket{\phi'_{B}}$.  The two qubits would pass their respective tests with
probability one.  Thus, even though Alice and Bob may not be able to exploit the cloned
state for some purpose, they can perform a measurement to confirm that the cloning has
taken place.  The no-cloning theorem does not simply prohibit {\em useful} cloning, 
in which both copies persist;
it forbids a two-qubit clone state at any stage of the evolution.  This theorem should
hold in any frame of reference in which quantum mechanics is valid.

Traditional arguments against superluminal communication generally proceed by constructing
causal paradoxes \cite{causality}.  
Our argument is based on quantum mechanics.  Indeed, the
only piece of the quantum theory that we have used is the linearity of the Schr\"{o}dinger
equation describing time evolution.  The principle of superposition is a very general
feature of quantum mechanics.

From time to time, it is suggested that quantum entanglement can be used to convey
classical messages instantaneously.  This cannot be done \cite{superlum}.  
We have here provided a new
elementary argument that exploits the properties of entanglement itself.  Far from 
providing a means of communicating instantaneously, quantum entanglement allows us to
exclude such a possiblity from a universe in which both quantum mechanics and relativity
hold true.

It has been pointed out that, if quantum cloning were possible, it could be used along
with entanglement to send superluminal classical signals \cite{no-cloning}.  
Our result is the converse:
since cloning is impossible, quantum entanglement can be used to show that superluminal
classical communication is impossible as well.

\section*{Acknowledgments}

A decade ago, John A. Wheeler posed the following question to one of the authors (B.S.): ``Can we find an argument against using entanglement for superluminal communication that is as simple and clear as the no-cloning theorem?'' This paper is the belated answer.

The authors wish to thank W. K. Wootters and S. Van Enk for valuable discussions.  
A note by V. S. Mashkevich \cite{mashkevich} was helpful in refining our argument.

\section*{Figure caption}

Spacetime diagrams illustrating how superluminal classical communication can violate
the quantum no-cloning theorem.
(a)  Teleportation of an unknown quantum state in the ``unprimed'' frame of reference.
If the classical information is sent superluminally from Alice to Bob, events I and II
have a spacelike separation.  (b)  The same teleportation process in the ``primed'' frame,
in which event II precedes event I.  At the ``$t' = $ constant'' hypersurface, both
qubits $C$ and $B$ are in the same state, as can be verified by measurements that are
simultaneous in this frame.


\begin{thebibliography}{9}
\bibitem{superlum}  A. Peres, {\em Quantum Theory:  Concepts and Methods},
	Kluwer Academic Publishers, Dordrecht (1993), p. 170.
\bibitem{no-cloning}  W. K. Wootters and W. H. Zurek, {\em Nature}
	{\bf 299}, 802 (1982).
\bibitem{teleportation}  C. H. Bennett, G. Brassard,
	C. Cr\'{e}peau, R. Jozsa, A. Peres, and W. K. Wootters, 
	{\em Physical Review Letters} {\bf 69}, 2881 (1992).
\bibitem{sendentang}  B. W. Schumacher, {\em Physical Review A} {\bf 54}, 2614 (1996).
\bibitem{relativity}  E. F. Taylor and J. A. Wheeler, {\em Spacetime Physics},
	W. H. Freeman and Company, San Francisco (1966), p. 62.
\bibitem{causality}  R. Mills, {\em Space, Time and Quanta}, 
	W. H. Freeman and Company, New York (1994), p. 64.
\bibitem{mashkevich}  V. S. Mashkevich, preprint quant-ph/9801062.
\end{thebibliography}
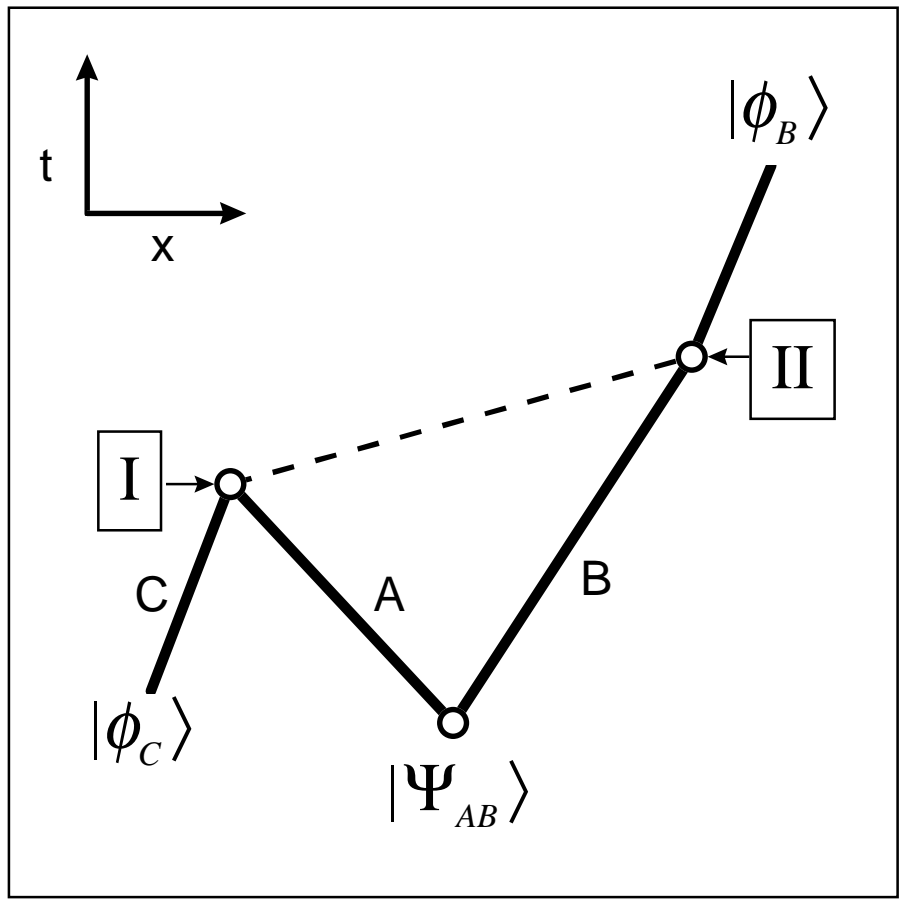
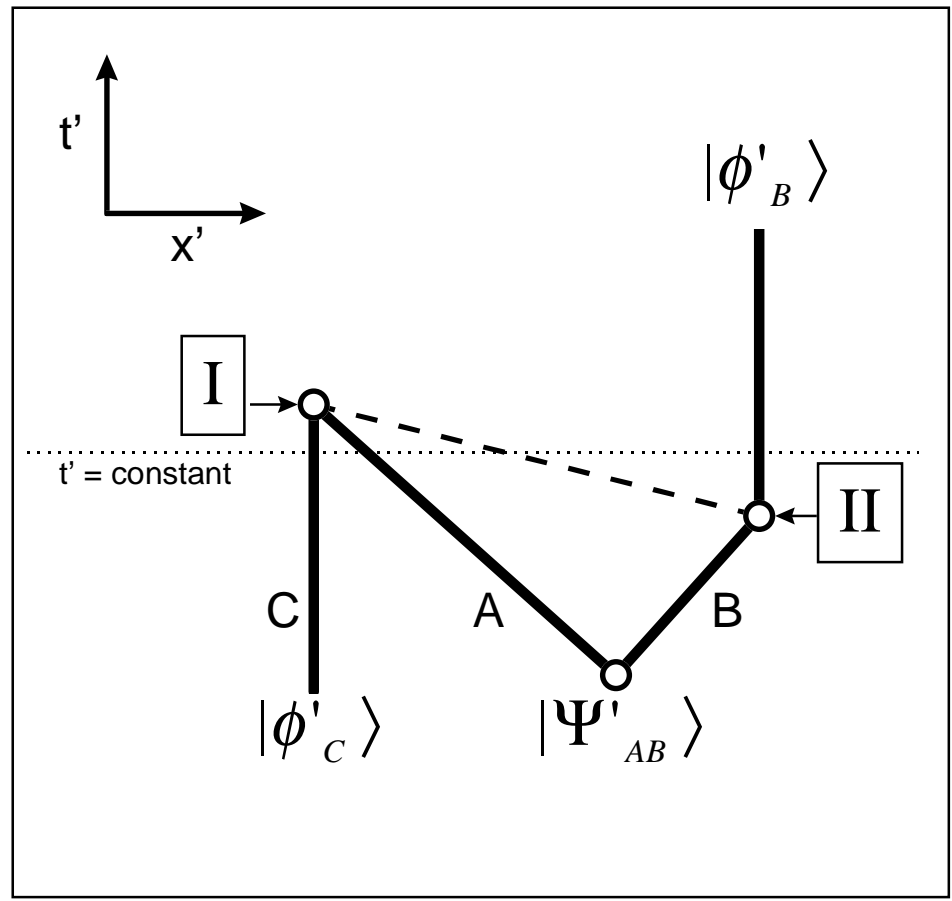
\end{document}